\documentclass[twocolumn,prl,aps]{revtex4}
\usepackage{amssymb,amsmath,amsthm}
\usepackage{graphicx}
\usepackage{verbatim}

\begin{document}
\title{Dissipation induced Instabilities and the Mechanical Laser}
\author{Marcel G. Clerc, Jerrold E. Marsden}
\address{Control and Dynamical Systems, 107-81, Caltech,
Pasadena, CA 91125}

\begin{abstract}
We study the 1:1 resonance for perturbed Hamiltonian systems with small
dissipative and energy injection terms. These perturbations of the 1:1
resonance exhibit dissipation induced instabilities. This mechanism allow us
to show that a slightly pumping optical cavity is unstable when one takes
into account the dissipative effects. The Maxwell-Bloch equations are the
asymptotic normal form that describe this instability when energy is
injected through forcing at zero frequency. We display a simple mechanical
system, close to the 1:1 resonance, which is a mechanical analog of the
Laser.
\end{abstract}

\maketitle

A fundamental problem in mechanics is the determination of the
stability of equilibria. Hamiltonian vector fields can undergo a
variety of instabilities as a single bifurcation parameter is
varied \cite{Abraham}. There are two fundamental codimension-one
bifurcations of Hamiltonian systems. The first is {\it the
stationary or steady state bifurcation}, which is characterized by
two eigenvalues merging at zero with multiplicity two
\cite{Marsden}. The second is {\it the 1:1 resonance}, which is
the collision of two pure imaginary eigenvalues (and their complex
conjugates) at finite frequencies with multiplicity two
\cite{LordKelvin}. This latter instability has been called by
different names depending on the field where it has appeared;  for
instance, `confusion of frequencies' in mechanical or electrical
oscillators \cite{Rocard}, `dispersive instability' \cite{Gibbon}
and `Hamiltonian Krein-Hopf bifurcation' \cite{Marsden}. These
instabilities are a consequence of the fact that in the
Hamiltonian case, if $\lambda $ is an eigenvalue, then
\cite{LordKelvin} so is $-\lambda $. The same property is present
in time reversible systems, i.e., they exhibit identical generic
instabilities. Recently, the instabilities of quasi-reversible
systems have also been characterized \cite{Clerc1999,Clerc1999b},
in which the irreversible effects are small and can be considered
as perturbative terms close to the instability. In particular,
systems close to the quasi-reversible 1:1 resonance are described
by the Maxwell-Bloch equations when the energy is injected through
a forcing at zero frequency \cite {Clerc1999b}. The Maxwell-Bloch
equations describe the interaction of an electromagnetic field and
a collection of two level atoms at an optical cavity \cite{Lamb}.

The aim of this letter is to study the 1:1 resonance for systems
in the neighborhood of a Hamiltonian one; that is, we shall
consider perturbed Hamiltonian systems with dissipative and energy
injection terms. Near this instability, the dissipative terms are
responsible for a spectral bifurcation, i.e., the dissipation
induces an instability \cite{Marsden1994}. This mechanism allows
us to show that a slightly pumping optical cavity is unstable when
one takes into account the dissipative effects. The Maxwell-Bloch
equations are the asymptotic normal form that describes this
instability in the presence of a conserved quantity.We shall
display a simple mechanical system, which we call the {\it
Mechanical laser}, which, close to the 1:1 resonance, is a
mechanical analog of the Laser.

For the sake of simplicity, we first consider a Hamiltonian system
with two degrees of freedom $H\left( p^{i},q_{i},\left\{ \lambda
\right\} \right) $, where $p^{i}$ and $q_{i}$ are canonically
conjugate variables, $\left( i=1,2\right) $, and $\left\{ \lambda
\right\} $ is a set of parameters. Assume there is a 1:1 resonance
for an equilibrium at $\lambda = \lambda _{c} $, i.e., the
spectrum at the equilibrium has a pair of pure imaginary
eigenvalues of multiplicity two, say at $\pm i\Omega $. Nearby,
the instability the system is governed by the normal form
\cite{Elphick}
\begin{eqnarray}
\partial _{t}A &=&i\Omega A+B  \nonumber \\
\partial _{t}B &=&i\Omega B+f\left( |A|^{2},i\left( AB^{\ast }-BA^{\ast
}\right) ,\left\{ \lambda -\lambda _{c}\right\} \right) A+  \nonumber \\
&&ig\left( |A|^{2},i\left( AB^{\ast }-BA^{\ast }\right) ,\left\{ \lambda
-\lambda _{c}\right\} \right) B,
\label{E-GlobalNormalForm}
\end{eqnarray}
where $f$ and $g$ are complex polynomial functions. There is a
change of variables from the given ones $\left\{
p^{1},q_{1},p^{2},q_{2}\right\}$ to the new ones (the complex
variables $A, B $) of the form $\left\{
p^{1},q_{1},p^{2},q_{2}\right\} =A\vec{\Psi}+B\vec{\chi}+
\vec{N}\left( A,B,A^{\ast },B^{\ast }\right) $, where $\vec{\Psi}$
is  the eigenvector of the linearized system corresponding to
$i\Omega $ and $\vec{\chi}$ is a generalized eigenvector in the
Jordan sense, and $\vec{N}\left( A,B,A^{\ast },B^{\ast }\right) $
is a nonlinear vector. This change of variables is not canonical
in general. When one considers rotated variables $\left(
A=e^{i\Omega t}A^{\prime },B=e^{i\Omega t}\partial _{t}A^{\prime
}\right) $ and the dominant terms, the normal form reads (omitting
the primes)
\begin{equation}
\partial _{tt}A=\varepsilon A+i\delta \partial _{t}A-\alpha |A|^{2}A
\label{E-reversalNormalForm}
\end{equation}
where $\varepsilon $ is the bifurcation parameter, which is
proportional to $ \lambda -\lambda _{c}$; henceforth we assume
$\varepsilon \ll 1$, $\delta $ is the gyroscopic term
\cite{LordKelvin} or detuning term \cite{Lamb} and $ \alpha $ is
an order one parameter. The variables and parameters scale as $
A\sim \varepsilon ^{1/2}$, $\partial _{t}A\sim \varepsilon $,
$\partial _{tt}A\sim \varepsilon ^{3/2}$, and $\delta \sim
\varepsilon ^{1/2}$. All terms in this equation are of order
$\varepsilon ^{3/2}$, while the higher order terms are
$\varepsilon ^{2}$.

When equation (\ref{E-GlobalNormalForm}) is written as a second
order system, it has an imaginary term linear in $A$, but such a
term is not present in (\ref{E-reversalNormalForm}) because it
breaks the eigenvalue symmetry ($\lambda \rightarrow -\lambda $).
The asymptotic  normal form (\ref {E-reversalNormalForm}) has the
following Hamiltonian
\begin{equation}
H=\partial _{t}A\partial _{t}A^{\ast }-\varepsilon
|A|^{2}+\frac{\alpha }{2} |A|^{4},  \label{E-Hamiltonian}
\end{equation}
with the Poisson-Bracket ($F$ and $G$ are real)
\[
\left\{ F,G\right\} =\frac{\partial F}{\partial A} \frac{\partial
G}{\partial A_{t}^{\ast }}-\frac{\partial G}{\partial
A}\frac{\partial F}{
\partial A_{t}^{\ast }}+i\delta \frac{\partial F}{\partial A_{t}}\frac{
\partial G}{ \partial A_{t}^{\ast }}+ {\rm c.c.}
\]
Thus, to cubic order, the nonlinear change of variables is
canonical. The eigenvalues of the zero solution ($A=0$) are $\pm
i\left( \delta /2\pm 1/2 \sqrt{\delta ^{2}-4\varepsilon }\right)
$. When $\varepsilon -\delta ^{2}/4$ is negative, the initial
Hamiltonian system has four distinct pure imaginary eigenvalues;
$\varepsilon -\delta ^{2}/4$ equal to zero is the 1:1 resonance at
$\pm \Omega $ frequencies and when it is positive, the eigenvalues
have nonzero real part. Note that the gyroscopic term is a
stabilizing effect \cite{LordKelvin}.

We now consider this Hamiltonian system under the influence of small
dissipative terms. This leads to a new term in the asymptotic normal form as
follows:
\begin{equation}
\partial _{tt}A=\varepsilon A-\left( \mu -i\delta \right) \partial
_{t}A-\alpha |A|^{2}A ,  \label{E-quasiReversalNormalForm}
\end{equation}
where $\mu $ is positive and order $\varepsilon ^{1/2}$.

To study the effects of dissipative terms in
(\ref{E-quasiReversalNormalForm} ), we consider its characteristic
polynomial, which has the form $\lambda ^{4}+2\mu \lambda
^{3}+\left( 2\varepsilon -\mu ^{2}+\delta ^{2}\right) \lambda
^{2}-\lambda 2\varepsilon \mu +\varepsilon ^{2}$. Looking for
roots of the form $\lambda =a\pm ib$ and $\lambda =c\pm id$, one
recognizes that $ \mu =-\left( a+c\right) $ and $\varepsilon \mu
=\left( c\left( a^{2}+b^{2}\right) +a\left( c^{2}+d^{2}\right)
\right) $. Hence, when $ \varepsilon $ is negative, $a$ and $c$
are negative, i.e., all eigenvalues are to the left of the
imaginary axis. For $\varepsilon $ positive and $ \varepsilon
-\delta ^{2}/4$ negative, the unperturbed system is marginal, but
the perturbed one satisfies $ac<0$; in addition, the eigenvalues
with larger frequency move to the left of the imaginary axis
(stable modes) and the others to the right (unstable modes), but
the eigenvalues that move furthest away from the axis are stable.
Finally, when $\varepsilon >\delta ^{2}/4$, the eigenvalues have
nonzero real part, and again the stable modes are the furthest
from the imaginary axis. Therefore, we are led to consider
dissipation induced instabilities \cite{Marsden1994}. The
destabilizing effects through positive or negative total
dissipative perturbation was known a long time ago by Lord Kelvin
et al. \cite{LordKelvin} From the physical point of view, one can
understand this phenomena as follows: when $ \varepsilon $ is
negative, the energy (\ref{E-Hamiltonian}) has a minimum at the
origin, hence when dissipation is added, the solutions near the
origin move towards it. By contrast, if $\varepsilon $ is
positive, the energy has a saddle point at the origin and when
$\varepsilon <\delta ^{2}/4$, this is unstable with an algebraic
evolution in time, so when one adds dissipation, this is
consistent with the conclusion that the solutions near the origin
move away from it, exponentially in time.

Using the preceding analysis, we infer that close to the 1:1
resonance, generically the dissipative terms induce an
instability. In the case that the instability happens with null
detuning ($\delta =0$), the normal form has only a real
coefficient and so the dissipative terms do not induce
instability. A physical example of this last situation is the
weakly dissipative baroclinic instability when the effect of
earth's sphericity is ignored \cite{Pedlosky}  and another is the
Kelvin-Helmholz instability \cite {Weissman}.

As we shall see in detail later, the Laser is a system that shows
dissipation induced instability, but firstly, we need to discuss
how the energy is injected to the system, so that the solution
that becomes unstable is persistent when the dissipative terms are
added. There are two natural ways to inject energy to the modes,
namely through forcing at finite frequency or at zero. The latter
situation is common in physical systems. In a Hamiltonian system
this is only possible if there is a conserved quantity, that is, a
zero eigenvalue whose mode is nonlinearly coupled with the other
ones. For instance, when there is a cyclic variable, the
respective momentum $Z$ is conserved. A Hamiltonian system that
has a 1:1 resonance in the presence of conserved quantity, leads
us to consider an extra equation of the form $\partial _{t}Z=0$
and an additional term $-ZA$ in the equation (
\ref{E-quasiReversalNormalForm}). The system presents different
behaviors depending on the value of $\epsilon -Z$. When one
includes dissipative terms and forcing, the asymptotic normal form
reads
\begin{eqnarray}
\partial _{tt}A &=&\left( \varepsilon -Z\right) A-\left( \mu -i\delta
\right) \partial _{t}A-\alpha |A|^{2}A  \nonumber \\
\partial _{t}Z &=&\nu Z+\eta |A|^{2}  \label{E-MaxwellBlochNF}
\end{eqnarray}
where $Z\sim \varepsilon $, $\nu \sim \varepsilon ^{1/2}$ and
$\eta \sim \varepsilon ^{1/2}$. The term $\eta |A|^{2}$ permits a
non trivial coupling between the variables. When the unperturbed
Hamiltonian system has more modes without resonances between them,
the perturbed system is governed by the above equations, since the
intensities of the other modes decreases in time. Through a
nonlinear change of variables, the previous equations are
equivalent to the Maxwell-Bloch equations \cite{Clerc1999b}. Using
a multiscaling method, the dispersive instability with small
dissipation is also described by the previous equations
\cite{Gibbon}.

To illustrate how dissipation induced instabilities enter, we
consider the semiclassical description of the Laser. This is based
on the self-consistent interaction of the electromagnetic field
with an active medium within an optical cavity. The electric field
is described classically (by the Maxwell equations) and the matter
as ensemble of atoms possessing two quantized energy levels;
phenomenological terms are added to complete the description.
Thus, the system is described by \cite{Lamb,Seigman}
\begin{align}
\frac{\partial ^{2}E}{\partial t^{2}}& =\frac{\partial
^{2}E}{\partial x^{2}} -\frac{\partial ^{2}P}{\partial
t^{2}}-\kappa \frac{\partial E}{\partial t},
\nonumber \\
\frac{\partial ^{2}P}{\partial t^{2}}& =-\gamma _{\perp
}\frac{\partial P}{
\partial t}-(\gamma _{\perp }^{2}+(1+\delta )^{2})P-gNE,  \nonumber \\
\frac{\partial N}{\partial t}& =-\gamma _{\parallel
}(N-N_{0})+E\left( \frac{
\partial P}{\partial t}+\gamma _{\perp }P\right) ,
\label{E-SemiClassical}
\end{align}
with periodic boundary condition at the cavity length ($L$,
$E\left( 0,t\right) =E\left( L,t\right) $). Here $E$,$\;P$ and $N$
are dimensionless quantities, which correspond to linearly
polarized electric field, the dipole polarization field and the
population inversion. $\gamma _{\parallel } $, $\gamma _{\perp }$
are the decay rate associated to spontaneous emission and
interaction between the atoms, $\kappa $ is a damping related to
the mirror losses, $\delta $ is the detuning, $g$ is a coupling
constant  which characterizes the atoms and $N_{0}$ the pump
parameter. In the time reversible limit of the above equations,
i.e., $\gamma _{\perp }=\gamma _{\parallel }=\kappa =0$, the
system has Hamiltonian density \cite{Holm}
\[
{\cal H}=\left( \frac{D}{4}-\frac{1}{\omega }Re\left(
a_{+}a_{-}^{\ast }\right) \right) ^{2}+\frac{\left( \partial
_{x}A\right) ^{2}}{2} +|a_{+}|^{2}-|a_{-}|^{2},
\]
with the Poisson-bracket
\begin{eqnarray*}
\left\{ F,G\right\}  &=&\int \frac{dx}{2}\left( \frac{\partial
F}{\partial A} \frac{\partial G}{\partial D}-\frac{\partial
G}{\partial D}\frac{\partial F}{
\partial A}\right) + \\
&&i\frac{\omega }{2}\left( \frac{\partial F}{\partial
a_{+}}\frac{\partial G }{\partial a_{+}^{\ast }}+\frac{\partial
F}{\partial a_{-}}\frac{\partial G}{
\partial a_{-}^{\ast }}\right) +{\rm c.c.}
\end{eqnarray*}
Where $N\equiv |a_{+}|^{2}-|a_{-}|^{2}$, $P\equiv \frac{4}{\omega
} \mathop{\rm Re}\left( a_{+}a_{-}^{\ast }\right) $, $E\equiv
A_{t}$, and $ \omega =1+\delta $. The Hamiltonian is just the sum
of the electromagnetic energy and the atomic excitation energy
\cite{Holm}.

Changing the cavity length in the time reversible limit of
equation (\ref {E-SemiClassical}), leads to a 1:1 resonance for
the non-lasing solution ($ E=P=0,$ $N=D_{0}$), which gives rise to
an electromagnetic wave with $\pm \omega $ frequencies. Using the
slowly varying envelope (WKB) approximation leads to the
Maxwell-Bloch equations \cite{Lamb}.

Figure 1 shows the space-time diagram of the electric field of a
numerical simulation of the semiclassical model, plus noise with
small intensity, close to the 1:1 resonance without dissipative
terms (see fig. 1a) and with dissipative ones (see Fig. 1b). The
numerical simulations start with the same initial condition,
namely, the no-lasing solution with excited atoms ($ D_{0}>0$). It
is clear from these pictures that the inclusion of dissipation
induces the Laser to respond. One can physically understand what
happens, since without dissipation the atoms excited decay through
the stimulated emission, i.e., the non-lasing solution becomes
unstable very slowly (nonlinear mechanism). Instead, when one
takes into account the dissipative terms, the excited atoms decay,
for instance through stimulated emission and collisions, i.e.,
exponentially in time. Note that, one must to pump to $
N_{0}=D_{0}$ in order that the non-lasing solution persists. In
brief, with nonzero detuning and slight pumping, the dissipation
induces the Laser to respond.

To illustrate the 1:1 resonance in a simple Hamiltonian system, we
consider a mechanical system, that we call the {\it Mechanical
Laser}, which consists of two coupled spherical pendula in a
gravitational field, with a support, which can rotate around a
vertical axis. The lower pendulum  is constrained to move in a
plane that is orthogonal to the plane of the upper pendulum (see
Fig. 2).

The system rotates with angular velocity $\dot{\varphi}$ with
respect to the vertical. The quantities $m_{1}$, $m_{2}$, $l_{1}$
and $l_{2}$ are the mass and length of the upper and lower pendula
respectively, and $I$ is the dimensionless moment of inertia of
the support. The system will dissipate energy because of friction
at the contacts and the motion of the pendulum masses in a fluid
(for example, the air) via Stokes' law. Energy is injected through
a constant torque at the upper pendulum pivot point. The governing
equations for the angles $\theta _{1}\left( t\right) $ and $\theta
_{2}\left( t\right) $ and the vertical angular velocity
$\dot{\varphi}$ read \cite{Clerc2001}
\begin{gather*}
\ddot{\theta}_{1}=-\sigma ^{2}\sin \theta _{1}\sin \theta
_{2}\ddot{\theta}
_{2}-\sigma ^{2}\sin \theta _{1}\cos \theta _{2}\dot{\theta}_{2}^{2} \\
-2\sigma ^{2}\cos \theta _{1}\cos \theta
_{2}\dot{\varphi}\dot{\theta}
_{2}+\sin \theta _{1}\cos \theta _{1}\dot{\varphi}^{2} \\
-\sigma ^{2}\cos \theta _{1}\sin \theta _{2}\ddot{\varphi}-\frac{g}{l}\sin
\theta _{1}-\nu _{1}\dot{\theta}_{1}, \\
\ddot{\theta}_{2}=-\sin \theta _{1}\sin \theta _{2}\ddot{\theta}_{1}-\cos
\theta _{1}\sin \theta _{2}\dot{\theta}_{1}^{2} \\
+2\cos \theta _{1}\cos \theta _{2}\dot{\varphi}\dot{\theta}_{1}+\sin \theta
_{2}\cos \theta _{2}\dot{\varphi}^{2} \\
+\sin \theta _{1}\cos \theta _{2}\ddot{\varphi}-\frac{g}{l}\sin \theta
_{2}-\nu _{2}\dot{\theta}_{2}, \\
\frac{d}{dt}\left\{
\begin{array}{c}
\left( \sin ^{2}\theta _{1}+\sigma ^{2}\sin ^{2}\theta _{2}\right)
\dot{
\varphi}+\sigma ^{2}\cos \theta _{1}\sin \theta _{2}\dot{\theta}_{1} \\
-\sigma ^{2}\sin \theta _{1}\cos \theta _{2}\dot{\theta}_{2}+I\dot{\varphi}
\end{array}
\right\}  \\
=-\nu _{\varphi }\left( \dot{\varphi}-\Omega \right) -\mu _{1}\sin
^{2}\theta _{1}\dot{\varphi}-\mu _{2}\left( \sin ^{2}\theta _{1}+\sin
^{2}\theta _{2}\right) \dot{\varphi}.
\end{gather*}
where $\nu _{1}$, $\nu _{2}$, $\nu _{\varphi }$, $\mu _{1}$ and
$\mu _{2}$ are damping coefficients, $\sigma =\sqrt{m_{2}/\left(
m_{1}+m_{2}\right) }$ is the relative factor of the energy between
the oscillators, and we have written the torque as $\nu _{\varphi
}\Omega $. For the sake of simplicity, we have considered the case
of pendula of equal lengths ($l_{1}=l_{2}=l$). When one regards
the Hamiltonian limit of the previous equations, the vertical
solution or non-lasing solution $\theta _{1}=\theta _{2}=0$,
$\dot{ \varphi}=\Omega _{0}$ has a 1:1 resonance when
\[
\Omega _{0}=\Omega _{c}=\sqrt{\frac{g}{l}\frac{\left(
m_{1}+m_{2}\right) }{ m_{1}}}
\]
with frequency $\omega _{c}=\pm \sqrt{gm_{2}/lm_{1}}$. The
centripetal force is more intense than the gravitational force
when $\Omega _{0}\geq \Omega _{g}\equiv \sqrt{g/l}$. As a
consequence, the Coriolis force exerted by one pendulum on the
other, the non-lasing solution is marginal when $\Omega _{g}\leq
\Omega _{0}\leq \Omega _{c}$. In this region  the system is
nonlinearly unstable and the system becomes linearly unstable when
$\Omega _{0}>\Omega _{c}$; this exhibits a coherence oscillation,
which is the signatures of the {\it laser instability}.

Near the 1:1 resonance, the coupled pendulum is govern by (\ref
{E-MaxwellBlochNF}), where\cite{Clerc2001}
\begin{eqnarray*}
\varepsilon  &=&2\frac{g}{l}\frac{\left( \Omega -\Omega _{c}\right) }{\Omega
_{c}},\text{\ }\alpha =\frac{g}{4l}\left( \frac{\sigma ^{4}-2\sigma
^{3}-2\sigma ^{2}+3}{1-\sigma ^{2}}\right) , \\
\delta  &=&2\sigma \left( \Omega -\Omega _{c}\right) ,\text{\ }\nu
=\frac{ \nu _{\varphi }}{I},\text{ \ }\mu =\frac{1}{2l^{2}}\left(
\nu _{1}+\nu
_{2}\right) , \\
\eta  &=&\frac{1}{I}\left[ \mu _{1}+\left( 1+\frac{1}{\sigma ^{2}}\right)
\mu _{2}-2\frac{\nu _{\varphi }}{\sigma ^{2}I}\left( \frac{\Omega }{\Omega
_{c}}-1\right) \right] ;
\end{eqnarray*}
where the variables are related to the dominate order by $A=\left(
\sigma \theta _{1}+i\theta _{2}\right) \exp (i\omega _{c}t)/2$ and
$Z=2\Omega _{c}\sigma ^{2}\left( \Omega _{c}-\dot{\varphi}\right)
$. Transforming equations (\ref{E-MaxwellBlochNF}) to the
Maxwell-Bloch equations, one obtains the following analogy of the
electric field, polarization and population inversion
\begin{eqnarray*}
E &=&\frac{\left\{ \sigma \theta _{1}+i\theta _{2}\right\} }{2}e^{i\left(
\Delta +\omega _{c}\right) t}+c.c. \\
P &=&\frac{\left\{ \sigma \dot{\theta}_{1}+i\dot{\theta}_{2}+a\left( \sigma
\theta _{1}+i\theta _{2}\right) \right\} }{2}e^{i\left( \Delta +\omega
_{c}\right) t}+c.c. \\
N &=&\frac{2\Omega _{c}\sigma ^{2}}{\alpha }\left( \dot{\varphi}-\Omega
_{c}\right) +\varepsilon _{0}-\frac{\sigma ^{2}\theta _{1}^{2}+\theta
_{2}^{2}}{4}
\end{eqnarray*}
where $\Delta =\delta \left( \nu \alpha +\eta \right) /\left( 2\nu
\alpha +2\eta -\mu \alpha \right) ,$ $a=\left( \nu \alpha +\eta
\right) /\alpha +i2\Delta $ and $\varepsilon _{0}=\left(
\varepsilon +\Delta ^{2}-\delta \Delta \right) /\alpha +\left( \nu
\alpha +\eta \right) (\mu \alpha -\nu \alpha -\eta )/\alpha ^{3}.$
The mechanical analog of the electric field is a simple function
of the pendulum displacements with respect to the vertical. In the
Hamiltonian case the polarization is the slow time derivative of
the electric field. The mechanical analogy of the  stimulated
emission mechanism (time reversal effect) is the vertical angular
momentum conservation. Thus, when one increases the pendula tilt
(the intensity of the electric field enlargement), the angular
velocity decreases (the population inversion declines). This is a
main ingredient of Laser theory. If the torque or pumping is
bigger than the gravitational force ($\Omega _{0}>\Omega _{g}$),
i.e., the system is excited in Laser terminology, then the torque
that gives rise to an angular velocity slightly higher than $
\Omega _{g}$ leads to a dissipation induced instability.

In summary, we have studied the 1:1 resonance for perturbed Hamiltonian
systems with small dissipative and energy injection terms. Nearby, the 1:1
resonance exhibits dissipation induces instabilities. This allow us to show
that a slightly pumping optical cavity is unstable when one takes into
account the dissipative effects. The Maxwell-Bloch equations are the
asymptotic normal form that describe this instability when energy is
injected through forcing at zero frequency. We have displayed a simple
mechanical system, the Mechanical laser or double spherical pendulum with a
support, which, close to the 1:1 resonance, is a mechanical analog of the
Laser.

\smallskip

We thank Richard Murray for his encouragement and help in making
this work possible and CDS for its hospitality. JEM was partially
supported by the National Science Foundation.

\begin{figure}[ht]
\includegraphics[scale=0.32,angle=0]{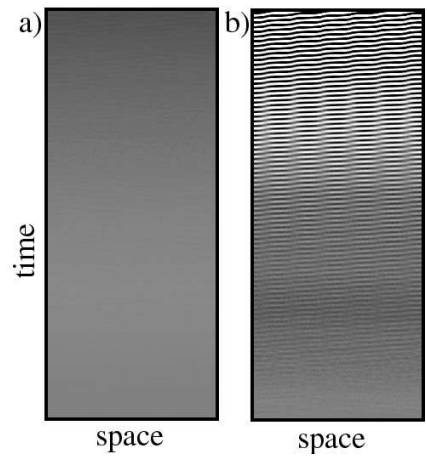}
\caption{{\protect\footnotesize Spatio-temporal\ diagram of the
electric
field of the semiclassical model (equation (\ref{E-SemiClassical})) with 
$L=120$, $\protect\delta =0.94$, $g=0.6$, $D_{o}=0.0338$ a) without
dissipation $\protect\kappa =\protect\gamma _{\perp }=\protect\gamma
_{\parallel }=0$, b) with dissipation $\protect\kappa =0.03$, 
$\protect\gamma _{\perp }=0.01 $, $\protect\gamma _{\parallel }=0.03$
 and $N_{o}=0.0338$.}}
\label{fig:spacetime_electric}
\end{figure}

\begin{figure}[ht]
\includegraphics[scale=0.5,angle=0]{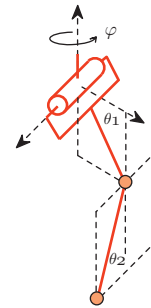}
\caption{{\protect\footnotesize Schematic representation of the
Mechanical Laser.}} \label{fig:mechanical_laser}
\end{figure}

\end{document}